\documentclass[prb,twocolumn,floatfix,showpacs]{revtex4}
\usepackage{graphicx}
\usepackage{amsmath,amssymb,amsfonts,bm}
\DeclareMathOperator{\sgn}{sgn}
\begin{document}
\title{Spin-orbit coupling effects in one-dimensional ballistic quantum wires}
\author{J.E.~Birkholz and V.~Meden} \affiliation{Institut f\"ur 
Theoretische Physik, Universit\"at G\"ottingen, 
Friedrich-Hund-Platz 1, D-37077 G\"ottingen, Germany}

\begin{abstract}
We study the spin-dependent electronic transport through a 
one-dimensional ballistic 
quantum wire in the presence of Rashba spin-orbit interaction.
In particular, we consider the effect of the 
spin-orbit interaction resulting from the lateral confinement of 
the two-dimensional electron gas to the one-dimensional wire geometry.
We generalize a situation suggested earlier
[P.~Str\v{e}da and P.~S\v{e}ba, Phys.~Rev.~Lett.~{\bf 90}, 256601
(2003)] which allows for spin-polarized electron transport. 
As a result of the lateral confinement, the spin is rotated out of the 
plane of the two-dimensional system. We furthermore investigate the
spin-dependent 
transmission and the polarization of an electron current at a 
potential barrier. 
Finally, we construct a lattice model which shows similar low-energy physics. 
In the future, this lattice model will allow us to study how the
electron-electron interaction affects 
the transport properties of the present setup.
\end{abstract}
\maketitle

\section{Introduction} 

Spin-orbit coupling is a relativistic effect of 
order $\mathcal{O}(v^2/c^2)$, where $v$ is the electron velocity,
which follows directly from the Dirac equation. It is 
described by the Hamiltonian (for $\nabla\times\mathbf{E}=0$)
\begin{equation}
H_\mathrm{SO}=-\frac{e\hbar}{4m^2 c^2} \mbox{\boldmath{$\sigma$}} \cdot 
\left[\mathbf{E}\times\left(\mathbf{p}-\frac{e}{c}\mathbf{A}\right)\right]
\, , 
\label{H_SO}
\end{equation} 
where the electric field $\mathbf{E}=-\nabla V/e$ ($e<0$ is the
electron charge) is the gradient of the ambient potential. 
In the following, the correction $-eA/c$ to the canonical momentum is
abandoned. In order to confine electrons to nanostructure devices,
sharp potentials are necessary, which lead to nonnegligible
spin-orbit interaction (SOI), especially in systems with structural
inversion asymmetry like e.g.~semiconductor heterostructures. This
effect can be used to achieve control over 
the electron spin and leads to spin-dependent transport properties, such as
spin-polarized currents, even in systems without ferromagnetic leads. 

The emerging field of spintronics might result in an extensive use of the
spin degree of freedom for information processing.\cite{review,winkler}
In a two-dimensional electron gas (2DEG) obtained by a strong confinement in 
$z$-direction, the SOI is usually described by the so-called Rashba term 
\begin{equation}
\label{Rashbaham}
H_R=\frac{\displaystyle \hbar}{\displaystyle m} \alpha_z
\left(\sigma_x p_y- \sigma_y p_x \right) \, ,
\end{equation} 
contributing to the Hamiltonian of the electron
system.\cite{rashba1,winkler} Here the components of the electron
momentum operator are denoted by $p_i$, the Pauli matrices by
$\sigma_i$, and $\alpha_z\propto E_z$ is the SOI coupling
coefficient\cite{footnote1}
set by the
confining electric field. As discussed by Datta and Das,\cite{datta} a
further confinement of the 2DEG to a wire geometry allows for a
particular control over the spin, if $\alpha_z$ or the length of the
wire are varied. This insight led to extensive studies on the
transport properties of noninteracting electrons in quasi
one-dimensional (1D) quantum wires with
SOI.\cite{moroz,mireles,uli,streda,pereira,serra,zhang}  
In particular, the effect of subband
mixing\cite{moroz,mireles,uli,serra} and a magnetic field
perpendicular to the plane of the underlying 2DEG\cite{zhang} 
was investigated.

\begin{figure}[tb]
\begin{center}
\includegraphics[width=0.23\textwidth,clip]{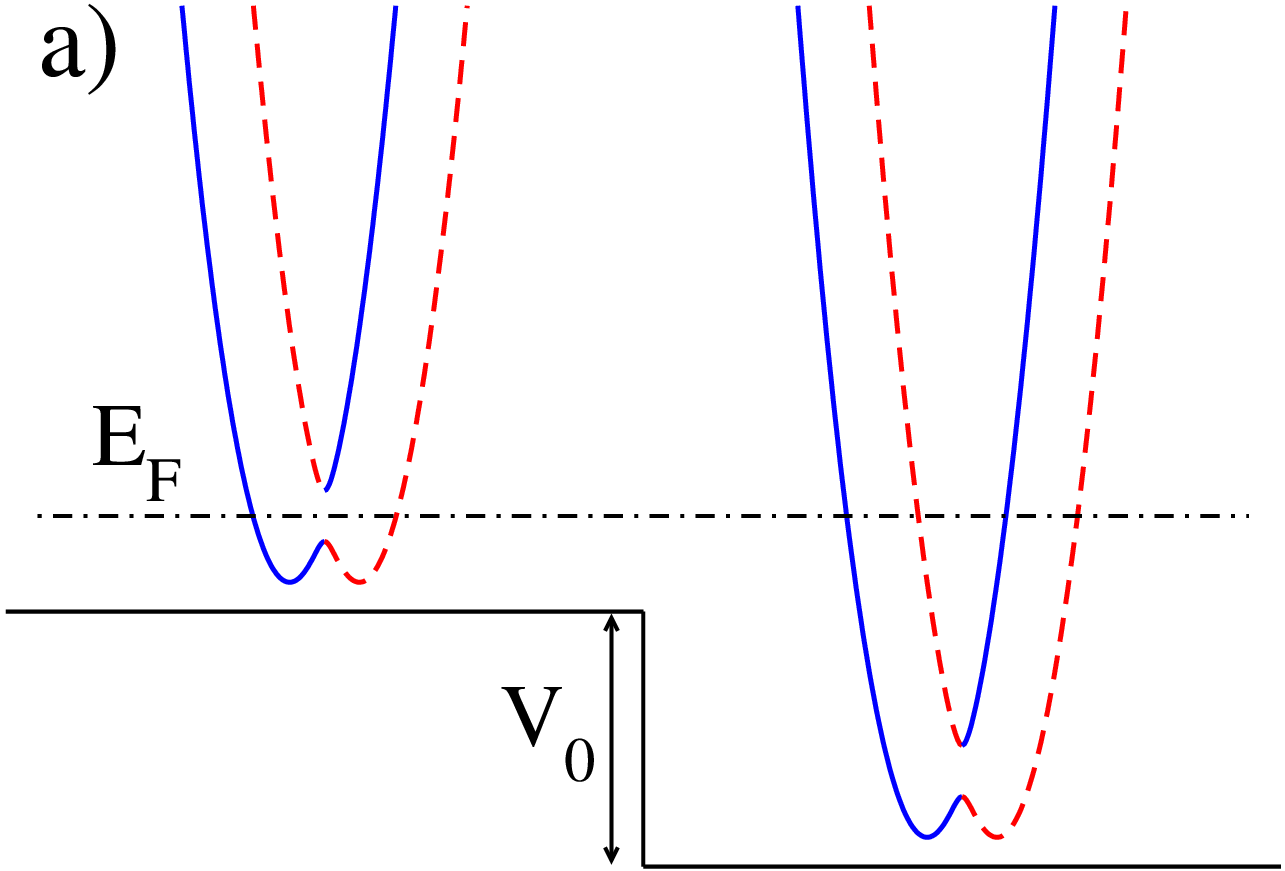}
\includegraphics[width=0.23\textwidth,clip]{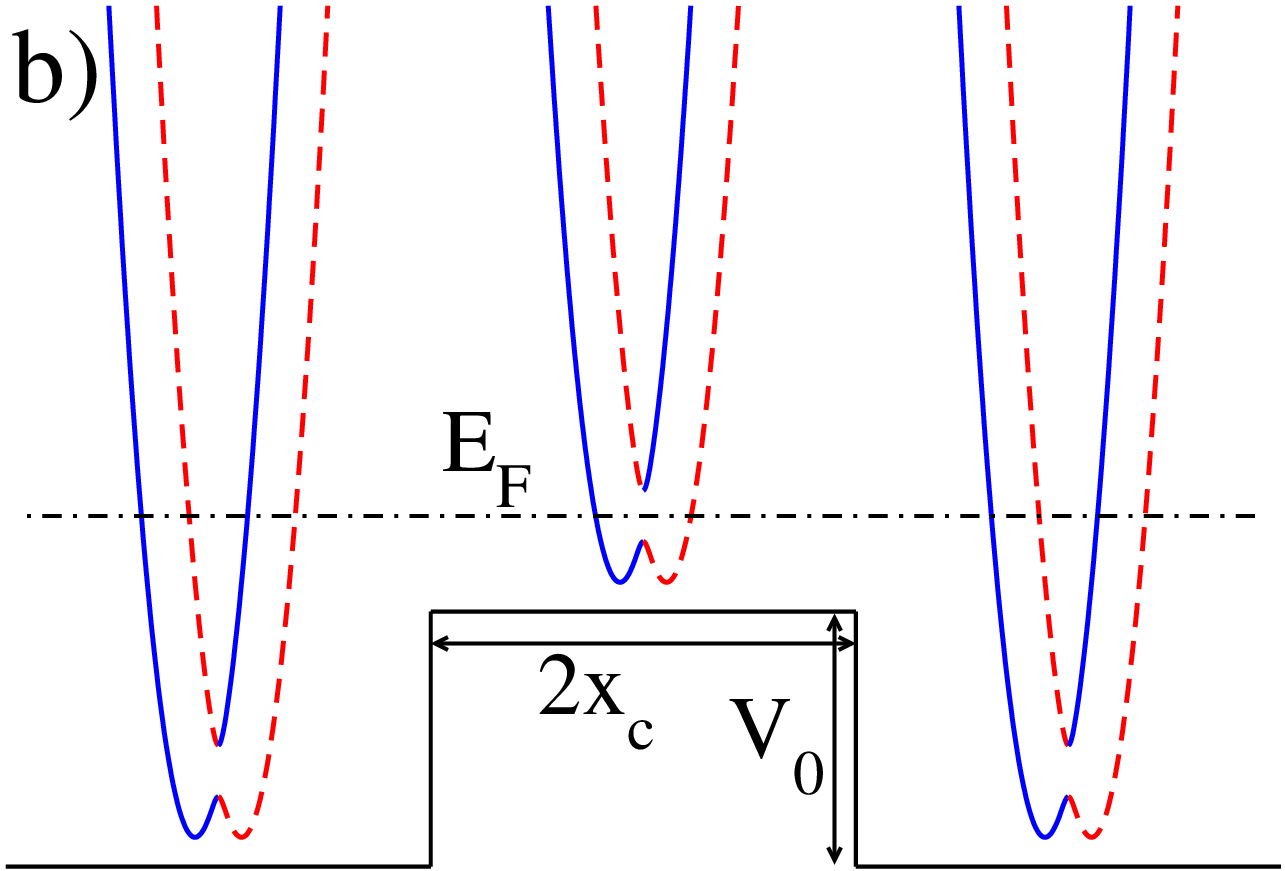}
\end{center}
\vspace{-0.6cm}
\caption[]{(Color online) a) A potential step of height $V_0$ and b) a potential barrier of height $V_0$ and width $2x_c$. The corresponding dispersions in the different regions are sketched (solid line: $s=+$, dashed line: $s=-$).}\label{potential}
\end{figure}

A very promising candidate for a system to experimentally produce
spin-polarized currents using SOI is the setup suggested by 
Str\v{e}da and S\v{e}ba 
where the magnetic field points in the wire direction and an
additional potential step is placed in the quantum
wire.\cite{streda,pereira}  It is assumed that due to the large energy 
level spacing only the lowest subband of the quantum wire is occupied
and subband mixing can be neglected. 
Restricting the considerations 
to this subband, one does not have to include explicitly the potential 
confining the electrons to the wire. 
Furthermore, the strong lateral confinement allows to take into account 
only the momentum in the wire direction, $p_x=p$, $p_y=p_z=0$ in 
Eq.~(\ref{Rashbaham}).
The energy dispersion of the 1D electron gas 
$\varepsilon_0(k) = \hbar^2 k^2/(2m)$, where $k=k_x$, is split by the 
Rashba term Eq.~(\ref{Rashbaham}) into two branches
$\varepsilon^{(s)}(k)=\hbar^2 (k +s \alpha_z)^2/(2m)-E_{\alpha_z}$, 
with $s= \pm$ and $E_{\alpha_z}=\hbar^2 \alpha_z^2/(2m)$.
The eigenenergies are fourfold degenerate with two left and 
two right moving states. The spin expectation values are 
$\langle \sigma_y \rangle_{k,s}=s$ and 
$\langle \sigma_x \rangle_{k,s}=\langle \sigma_z \rangle_{k,s}=0$, 
independent of $k$. In presence of an external magnetic field 
(parallel to the wire), described by a Zeeman term 
\begin{equation}
\label{Zeemanham}
H_Z= \epsilon_Z \sigma_x /2 \, , 
\end{equation}
an ``energy gap'' of size $\epsilon_Z$ opens up at $k=0$ 
[see Fig.~\ref{potential} a)]
and states within this ``gap'' are 
only twofold degenerate (one left and one right moving state).
A potential step can then be used to generate a tunable
spin polarization, in mainly the $y$-direction, of the linear response
current. In order to achieve this, the height of the step $V_0>0$ for wire positions
$x<0$ has to be chosen such that the energy falls into the ``gap'' region,
while for the potential free part $x>0$ it lies sufficiently above 
the ``gap'' [see Fig.~\ref{potential} a)].
As an additional effect of the magnetic field, the
spin expectation value is rotated gradually from the $\pm y$-direction 
into the $\pm x$-direction when $|k|\to 0$, while $\left< \sigma_z \right>_{k,s}$ 
remains zero. 
Depending on the chosen parameters, this leads to a small 
$x$-component of the ground state magnetization, whereas the $y$- and
$z$-components are exactly zero as will be explained below.

We here generalize the situation studied in Ref.~\onlinecite{streda} in
several ways.
We first study how the above scenario is modified in the presence of an
additional Rashba term 
\begin{equation}
\label{Rashbahamprime}
H_R'=\frac{\hbar}{m} \alpha_y \sigma_z p_x
\end{equation} 
resulting from the confinement of the 2DEG to the wire
geometry, a term which so far was mainly ignored. 
As we also focus on the lowest subband and do not study
subband mixing, the exact shape of the potential confining the electrons to 
the wire is not important.
As its main effect, $H_R'$ will lead to
nonvanishing spin expectation values $\langle \sigma_z \rangle_{k,s}$
and thus a spin polarization component perpendicular to the plain of
the underlying 2DEG. 
We also study the transmission current and the spin polarization at a potential barrier 
and discuss the interplay of $\alpha_y$ and $\alpha_z$. 
In addition, we present a lattice model which in an appropriate
parameter regime shows the same physics as the continuum model. This
model will allow us to study the effect of the electron-electron
interaction on the spin polarization in a forthcoming
publication\cite{interactionpaper} using the functional
renormalization group method.\cite{funRG} 

\section{Continuum model} 

\begin{figure*}[t]
\begin{center}
\includegraphics[width=0.3\textwidth,clip]{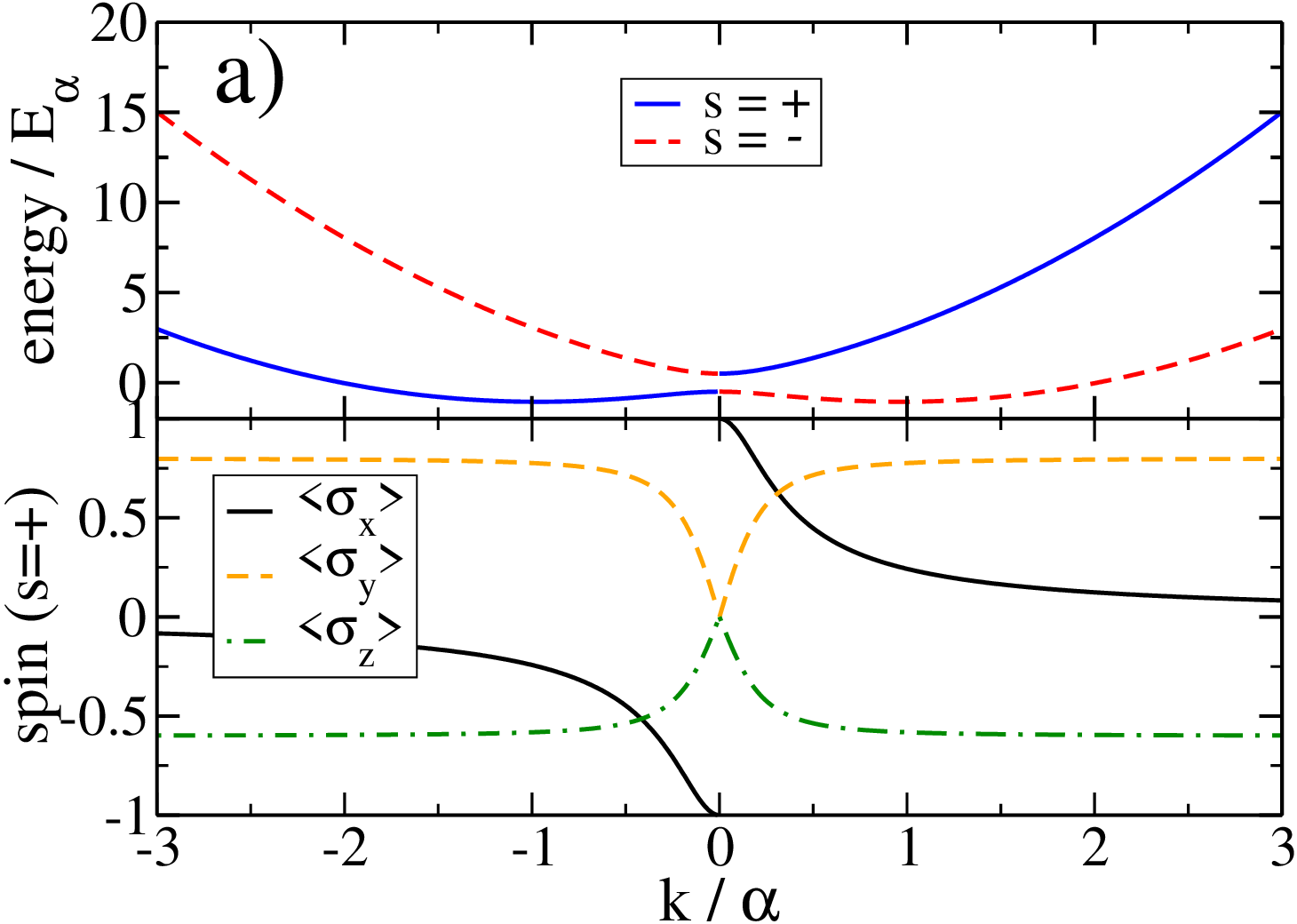}\hspace{0.5cm}
\includegraphics[width=0.3\textwidth,clip]{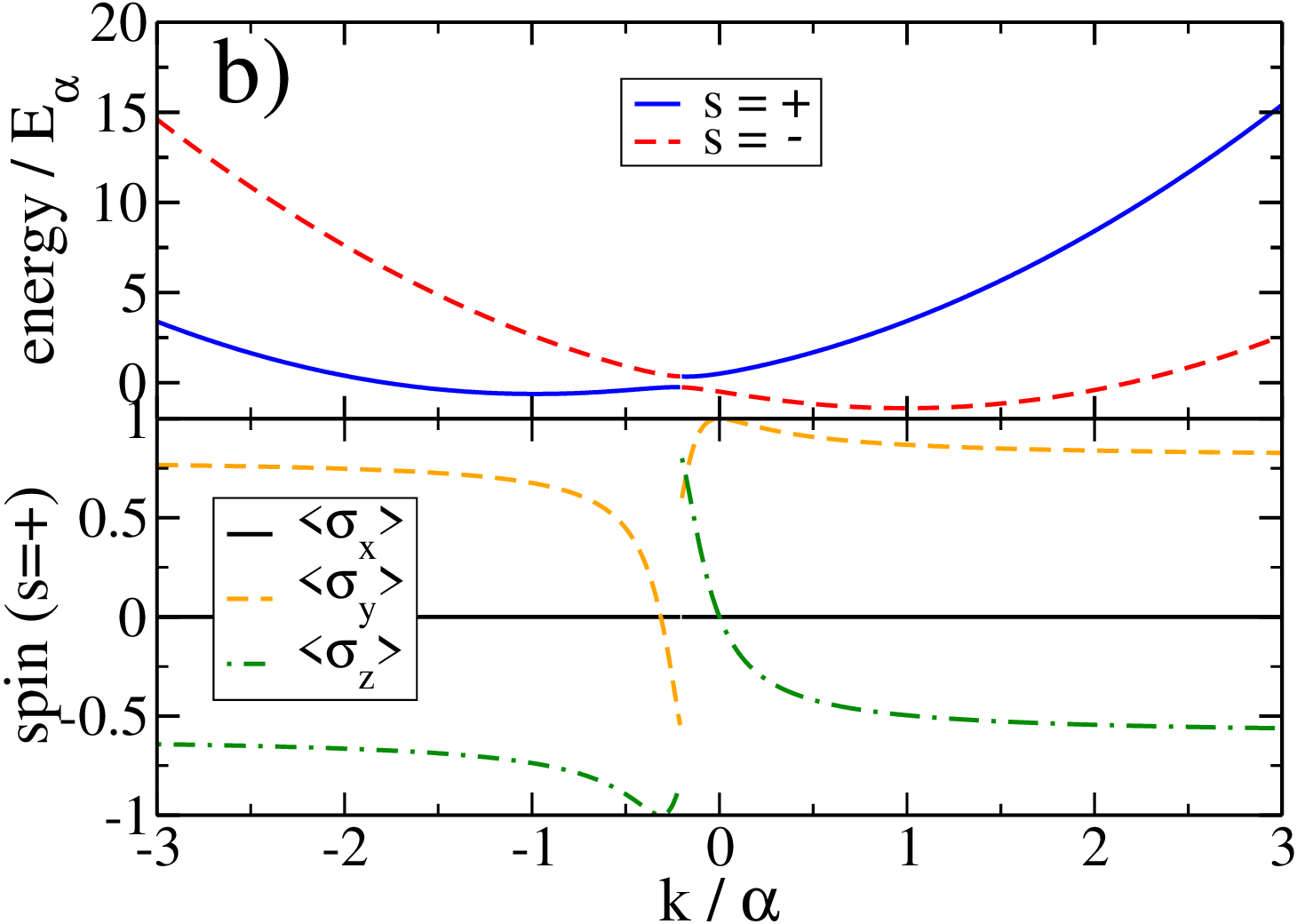}\hspace{0.5cm}
\includegraphics[width=0.3\textwidth,clip]{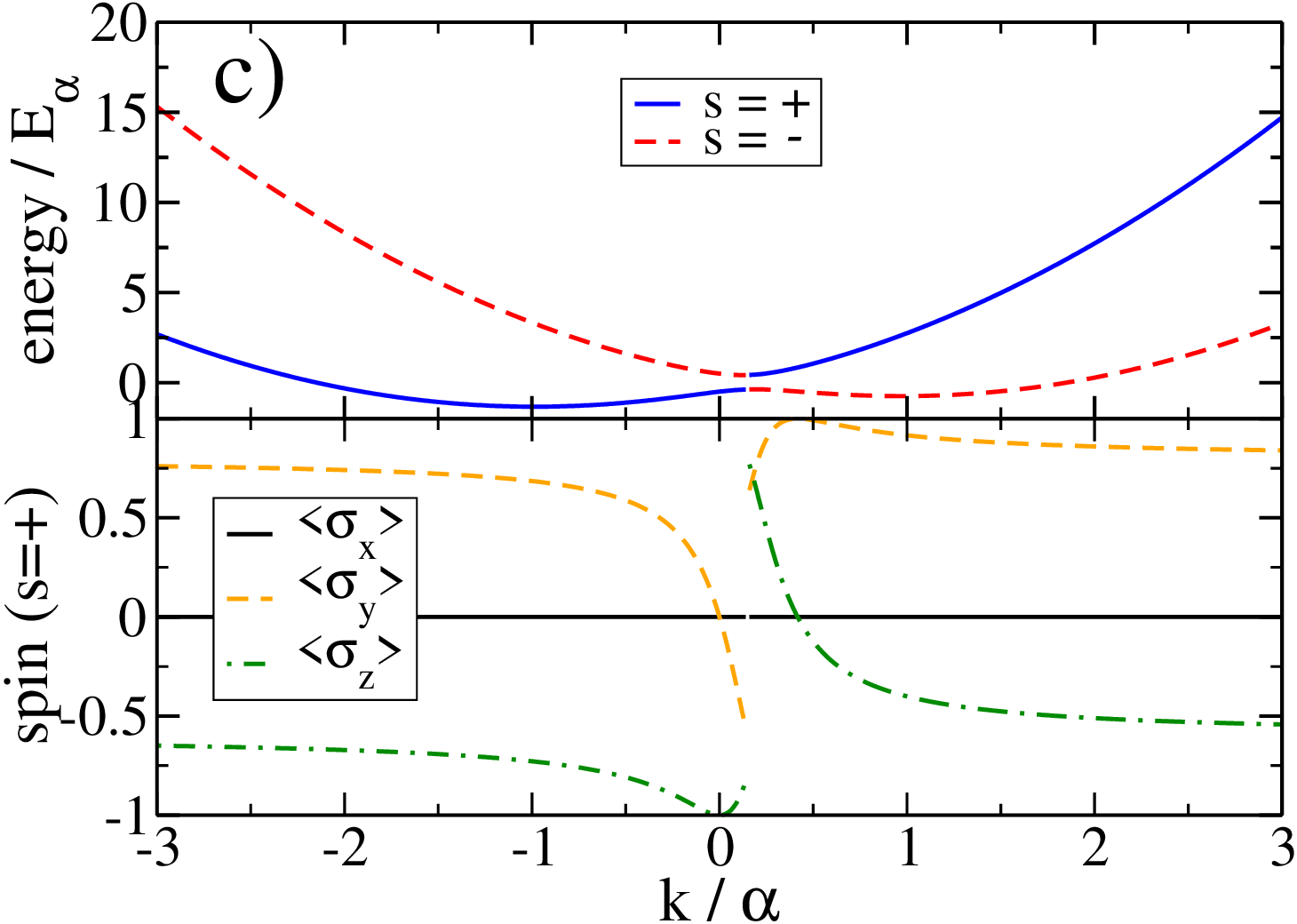}
\end{center}
\vspace{-0.6cm}
\caption[]{(Color online) Dispersion and spin expectation values 
  on the ($s=+$)-branch for a
  magnetic field in a) $x$- , b) $y$- and c) $z$-direction,
  $\alpha_y/\alpha=-0.6$, $\alpha_z/\alpha=-0.8$,
  $k_Z/\alpha=0.5$. The spin on the ($s=-$)-branch points
  in the opposite direction, i.e.~$\langle \sigma_i\rangle_{k,s}=
  -\langle \sigma_i\rangle_{k,-s}$. The shape of the dispersion and the
  $k$-value at which the ``energy gap'' becomes smallest clearly
  depends on the direction of the magnetic field.} 
\label{spindisp}
\end{figure*}

The model we consider is given by the Hamiltonian 
\begin{eqnarray}
\label{totham}
H=\frac{p_x^2}{2m} - \frac{\hbar\alpha_z}{m} \sigma_y p_x +
\frac{\hbar\alpha_y}{m} \sigma_z p_x
-\frac{e\hbar}{2mc}\mbox{\boldmath{$\sigma$}}
\cdot \mathbf{B} \; .
\end{eqnarray}
We slightly generalized the situation discussed above and allow for a
Zeeman term with a magnetic field $\mathbf{B} =
B (\sin{\theta}\cos{\varphi},\sin{\theta}\sin{\varphi},\cos{\theta})$
pointing in arbitrary direction. The normalized eigenstates with
quantum numbers $k$ and $s=\pm$ are given by the 
product of a plane wave (in $x$-direction) and a two-component spinor 
\begin{equation}
\label{ansatz}
\phi_k^{(s)}(x) =
\frac{\displaystyle{1}}{\displaystyle \sqrt{2\pi}}e^{ikx}\left(\begin{array}{c} 
A_k^{(s)}\\ B_k^{(s)}\end{array}\right) \,. 
\end{equation} 
Applying the Hamiltonian Eq.~(\ref{totham}) to this ansatz we obtain
\begin{widetext}
\begin{equation}
\left(\begin{array}{c c} k^2+2 \alpha_y k + 2 k_Z^2 \cos\theta -\epsilon, 
& 2ik\alpha_z+ 2 k_Z^2 e^{-i\varphi}\sin\theta \\
2ik\alpha_z+ 2k_Z^2 e^{i\varphi}\sin\theta, & k^2-2\alpha_y k - 2k_Z^2 \cos\theta 
-\epsilon \end{array} \right) \left( \begin{array}{c} A_k^{(s)}\\ 
B_k^{(s)} \end{array}\right)=0 \, ,
\end{equation}
with 
$\epsilon=2m E/\hbar^2$, $\alpha_y=eE_y/(4mc^2)$, $\alpha_z=eE_z/(4mc^2)$,
and $k_Z^2=-eB/(2\hbar c)$. 
Note that $\alpha_y,\alpha_z<0$ in our notation due to the negative electron charge. 
One obtains the eigenenergy (divided by $\hbar^2/2m$)
\begin{equation}
\epsilon^{(s)}(k)=k^2+2s\sgn (k-k_0) \sqrt{C(k)} \, , \label{eigenvalue}
\end{equation} 
with $C(k)=(\alpha_y^2+\alpha_z^2)k^2+2k_Z^2 k(\alpha_y\cos\theta - \alpha_z\sin\theta\sin\varphi) + k_Z^4$ and $k_0=-k_Z^2\left(\alpha_y\cos\theta-\alpha_z\sin\theta \sin\varphi\right)/(\alpha_y^2+\alpha_z^2)$ being the wave number at which the ``energy gap'' becomes smallest [see Fig.~\ref{spindisp}]. The corresponding eigenfunctions are
\begin{equation}
\phi_k^{(s)}(x)=\frac{1}{\sqrt{2\pi}\sqrt{1+\left|a_k^{(s)}\right|^2} }e^{ikx}\left(\begin{array}{c} 
a_k^{(s)}\\ 1 \end{array}\right) \, , \label{eigenfunction}
\end{equation}
\begin{equation}
\mbox{with}\quad a_k^{(s)}= \frac{-i\alpha_z k-k_Z^2 e^{-i\varphi}\sin\theta}{\alpha_y k + k_Z^2\cos\theta - s\sgn (k-k_0)\sqrt{(\alpha_y^2+\alpha_z^2)k^2+2k_Z^2 k(\alpha_y\cos\theta - \alpha_z\sin\theta\sin\varphi)+k_Z^4}}\label{aks}
\end{equation}
\end{widetext}
and the spin expectation values are given by
\begin{equation}
\langle \sigma_x + i\sigma_y\rangle_{k,s} =
2\frac{\left(a_k^{(s)}\right)^*}{1+\left|a_k^{(s)}\right|^2} \, , \quad
\langle \sigma_z \rangle_{k,s} =
\frac{-1+\left|a_k^{(s)}\right|^2}{1+\left|a_k^{(s)}\right|^2}\label{spin}
\, .
\end{equation}
As can be seen from Eq.~(\ref{spin}), the necessary condition
$\langle\sigma_x\rangle^2_{k,s}+ 
\langle\sigma_y\rangle^2_{k,s}+\langle\sigma_z\rangle^2_{k,s}=1$ holds for all
values of $s$ and $k$.
The existence of the confinement in $y$-direction (represented by
$\alpha_y$) leads to a rotation of the spin out of the $x$-$y$-plain
into the $z$-direction. This indicates that the ratio of $\alpha_y$
and $\alpha_z$ is crucial for the spin direction.
 
The energy dispersion Eq.~(\ref{eigenvalue}) and the spin expectation
values on the ($s=+$)-branch are shown in Fig.~\ref{spindisp} as a
function of $k$, with $k$ given in units of
$\alpha=\sqrt{\alpha_y^2+\alpha_z^2}$ and the energy in units of
$E_\alpha=\hbar^2\alpha^2/2m$. 
For $|k| \gtrsim \alpha$ the spin expectation values reach 
their asymptotic, $k$-independent values. 
The spin on the ($s=-$)-branch points
in the opposite direction, i.e.~$\langle
\sigma_i\rangle_{k,s}=-\langle \sigma_i\rangle_{k,-s}$, and is not
shown explicitly here. 
In combination with the fact
that for $\mathbf{B}=(B,0,0)$, $\langle \sigma_y\rangle_{k,s}$ and $\langle
\sigma_z\rangle_{k,s}$ are symmetric with respect to $k=0$ on both branches,
this explains 
why there is no ground state magnetization in the $y$- and $z$-direction
for $\mathbf{B}$ being parallel to the wire. However, there is a 
nonvanishing ground state magnetization in the $x$-direction.
The ``energy gap'' is given by $4\sqrt{C(k_0)}$ 
[see Eq.(\ref{eigenvalue})] and does not necessarily decrease from its 
maximum value $4k_Z^2$, if $\mathbf{B}$ is tilted against $\mathbf{e}_x$ as 
stated in Ref.~\onlinecite{streda}. 
In units of the Zeeman energy $E_Z=2\hbar^2 k_Z^2/2m$, the size
of the ``gap'' $E_G$ for arbitrary magnetic field 
$\mathbf{B}=B(\sin\theta\cos\phi,\sin\theta\sin\phi,\cos\theta)$ is given by
\begin{equation}
\frac{E_G}{E_Z}=1-\frac{\left(\alpha_y\cos\theta-\alpha_z\sin\theta\sin\phi\right)^2}{\alpha^2}.
\label{energygap}
\end{equation} 
Therefore, a finite $\alpha_y$ term is necessary for opening the ``gap'' for 
$\mathbf{B}||\mathbf{e}_y$. 
To emphasize this effect, we choose the parameter set
$(\alpha_y,\alpha_z,k_Z)/\alpha=(-0.6,-0.8,0.5)$ in Fig.~\ref{spindisp}.
In many experimental systems the confining potential in the $y$-direction might be 
much weaker than in the $z$-direction. In this case $|\alpha_y| \ll |\alpha_z|$ but 
subband mixing becomes relevant. The latter strongly affects the spin-dependent 
transport properties as e.g.~investigated in Ref.~\onlinecite{uli}, and the polarization 
effects discussed here can be expected to disappear. To achieve spin polarization in the 
present setup a strong confinement in the $y$-direction leading to a sizable 
$\alpha_y$ is thus essential. The lower dispersion branch in 
Fig.~\ref{spindisp} has a ``W''-like shape. For
$\mathbf{B}=(B,0,0)$, the condition for this behavior is
$\alpha_y^2+\alpha_z^2>2k_Z^2$ and becomes much more complex for
arbitrary magnetic field. We will focus on the situation where
$\mathbf{B}=(B,0,0)$.

The transmissions $t_{ss'}$ (conductance divided by $e^2/\hbar$) of an
electron current at fixed Fermi energy $E_F$ passing a potential
step in the wire direction [see Fig.~\ref{potential} a)] 
are obtained by assuming continuity of the wave functions and their 
derivatives at the interface. Here the first
index labels the branch to the left and the second index labels
the branch to the right of the potential step.
It was argued in
Ref.~\onlinecite{molenkamp} that one has  
to consider the continuity of the wave function's flux and not simply 
its derivative, but in our setup both conditions lead to the same
equations as we consider a homogeneous SOI.
The total transmission $T$ is the sum of
the four components $t_{++}$, $t_{+-}$, $t_{-+}$, and $t_{--}$. To the right of the
potential and for momenta $|k| \gtrsim \alpha$,
one can assign spins with quantum numbers
$\uparrow,\downarrow$ and a properly chosen quantization axis to the
branches $s=+,-$ because of the independence of
$\langle\mbox{\boldmath{$\sigma$}}\rangle_{k,s}$ on $k$.
However, the polarization vector is given by
\begin{equation}
\mathbf{P}=\frac{t_{++}+t_{-+}}{T}
\langle\mbox{\boldmath{$\sigma$}}\rangle_{k,+}+\frac{t_{+-}+t_{--}}{T}
\langle\mbox{\boldmath{$\sigma$}}\rangle_{k,-} \, . \label{polar}
\end{equation}

Since the potential step geometry was already 
discussed,\cite{streda} we will only shortly mention the influence of
the additional term $H_R'$, defined in Eq.~(\ref{Rashbahamprime}), and
discuss the interesting case
of a potential barrier [see Fig.~\ref{potential} b)] in more
detail. The latter can experimentally be achieved by adding gates to
the 1D quantum wire.

\begin{figure}[tb]
\begin{center}
\includegraphics[width=0.4\textwidth,clip]{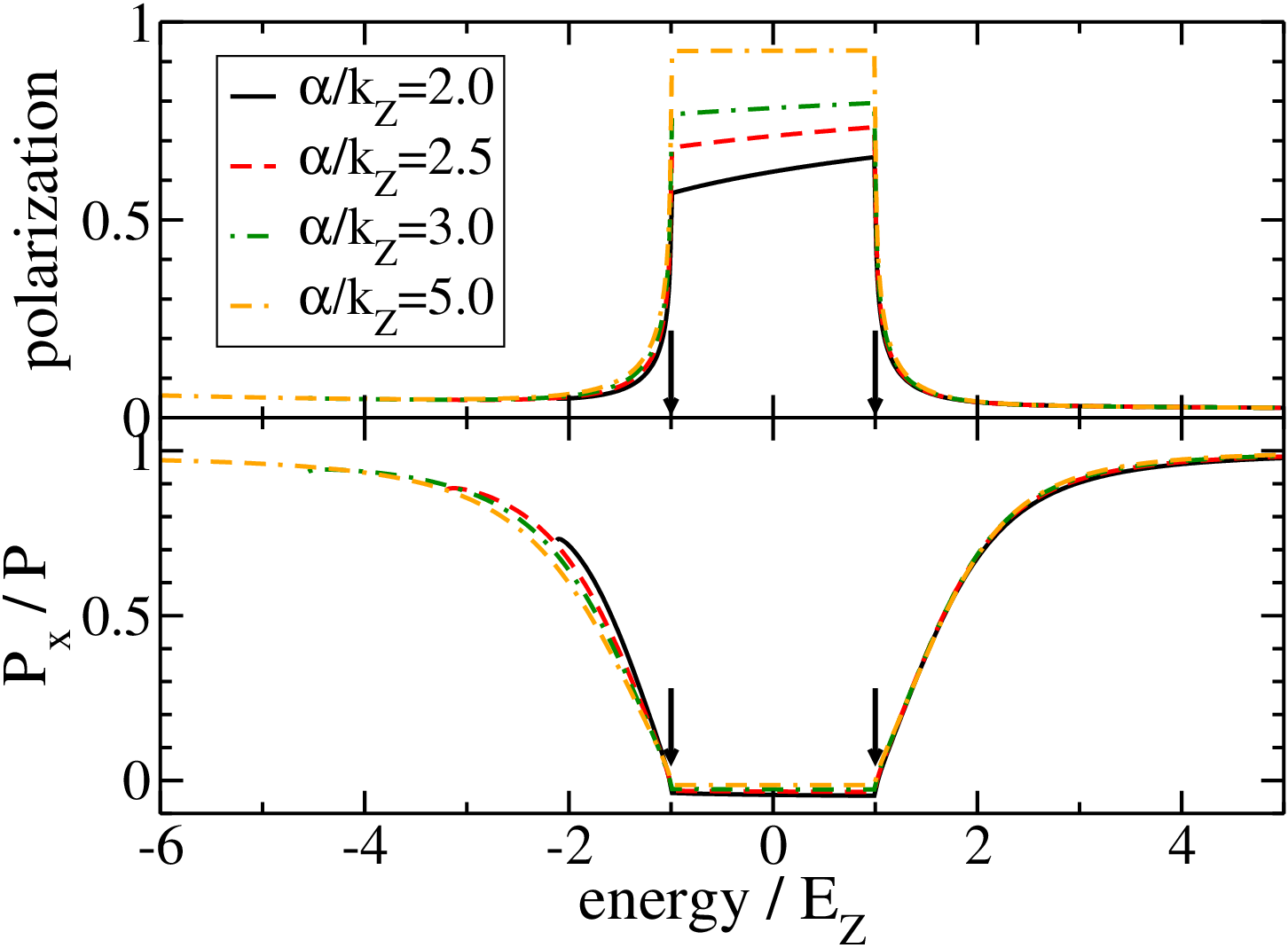}
\end{center}
 \caption[]{(Color online) Polarization of the transmission
   current at a potential step as a function of the Fermi energy for
   $V_0/E_Z=15$ and $\alpha/k_Z=2$, $2.5$, $3$, $5$. The total polarization $P$ 
   is sizable for
   energies in the ``gap'' (indicated by the arrows). In this regime it is
   mostly carried by $P_y$ and $P_z$. The polarization
   becomes negligible for energies outside the ``gap'' where $P_x$
   dominates.} 
\label{step_pol}
\end{figure}

\begin{figure}[tb]
\begin{center}
\includegraphics[width=0.4\textwidth,clip]{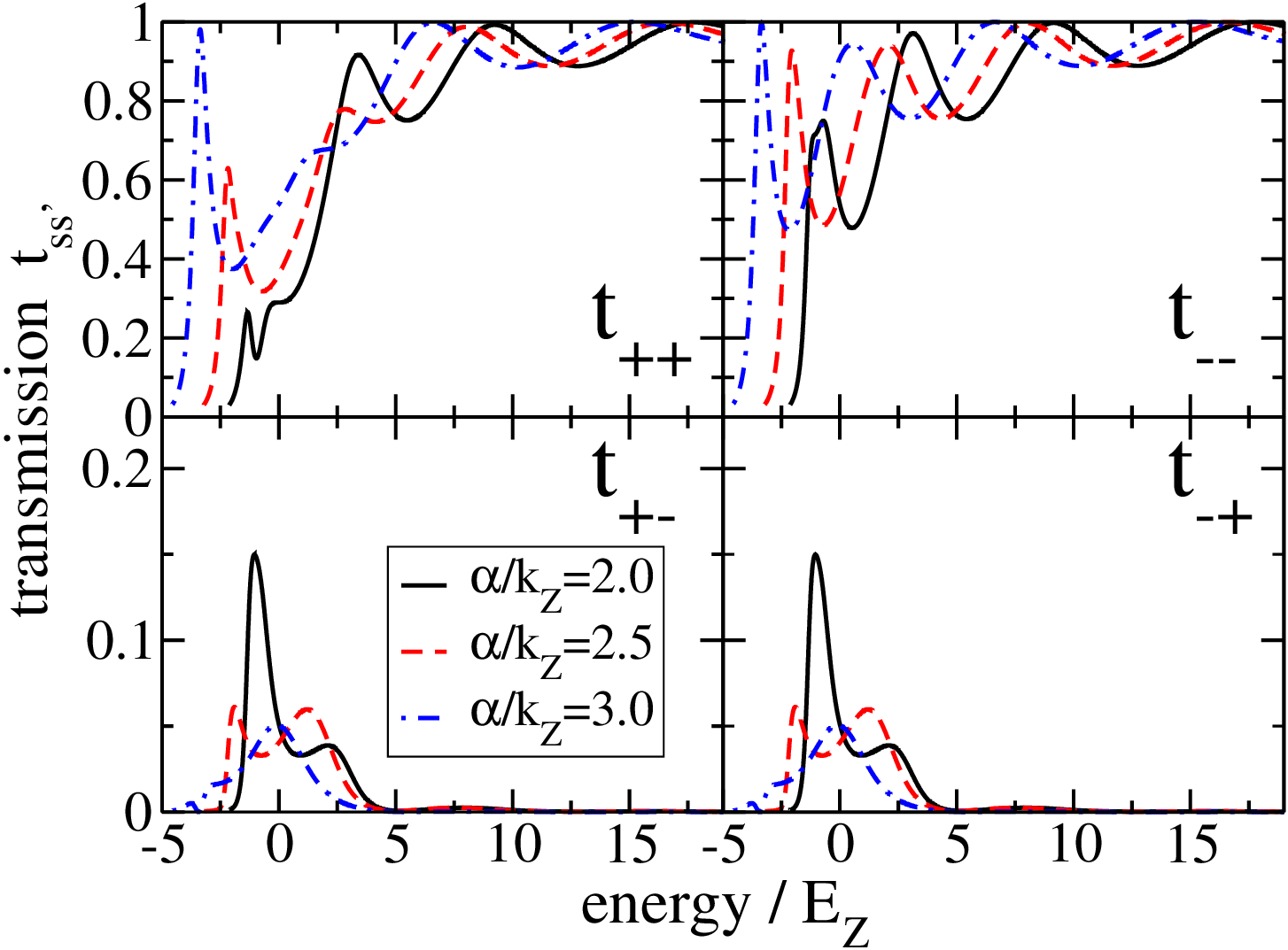}
\end{center}
\caption[]{(Color online) Partial transmissions at a potential
  barrier as a function of the energy for $V_0/E_Z=15$ and
  $\alpha/k_Z=2$, $2.5$, $3$. } 
\label{barrier_G}
\end{figure}

\begin{figure}[tb]
\begin{center}
\includegraphics[width=0.4\textwidth,clip]{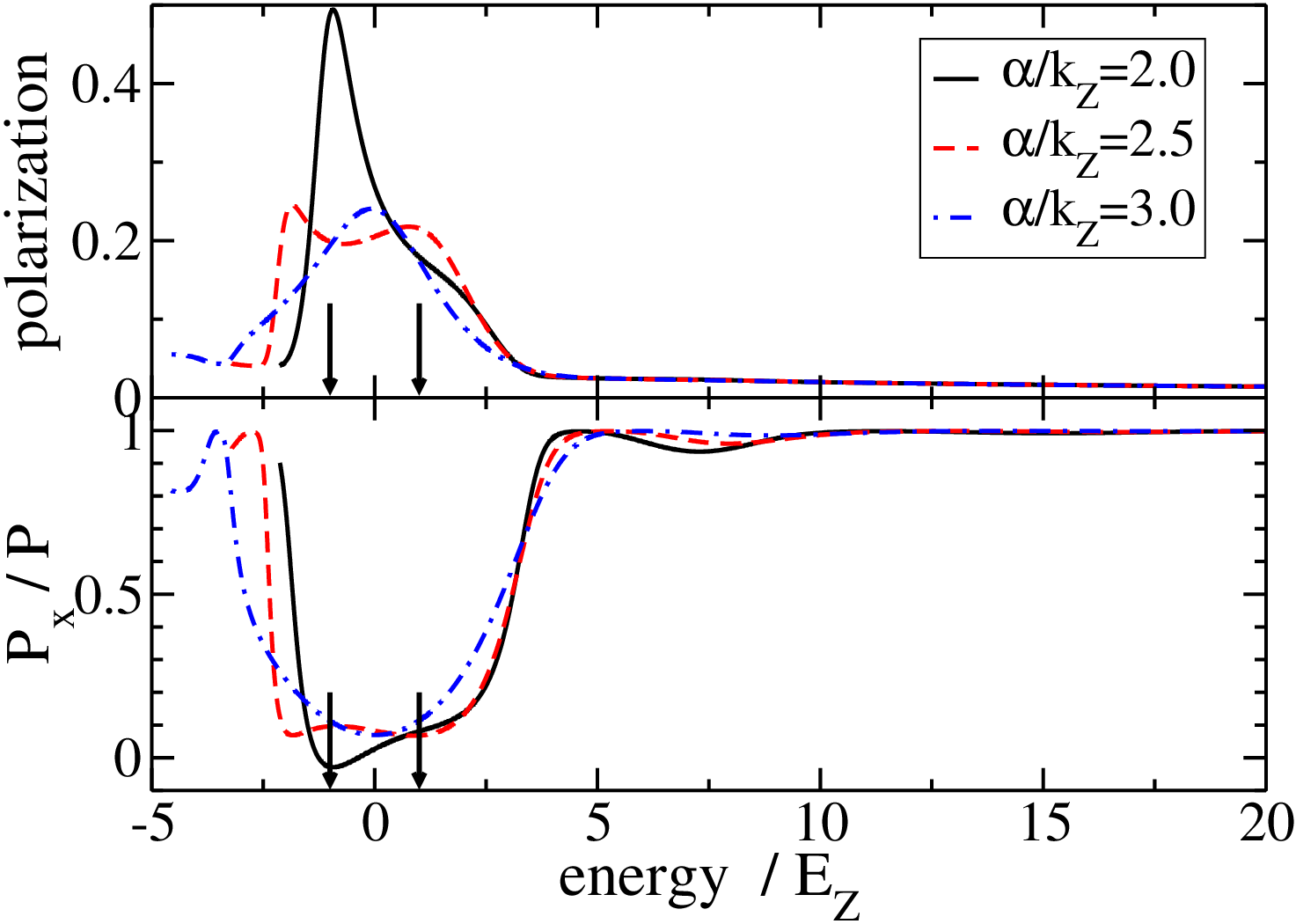}
\end{center}
\caption[]{(Color online) Polarization of the transmission
  current at a potential barrier as a function of the energy for the
  same parameters as in Fig.~\ref{barrier_G}. The polarization is sizable
  for energies well beyond the ``gap'' (indicated by the arrows) and
  shows oscillatory behavior. The $x$-component $P_x$ is only relevant 
  in regimes where the total polarization is small.} 
\label{barrier_P}
\end{figure}

As shown in Fig.~\ref{step_pol}, the total polarization
$P= | \mathbf{P} |$
of the current passing the potential step is large for energies in the
``gap'' and increases with
$\alpha$. Similar to the transmissions $t_{ss'}$, $P$ as well as 
the parallel polarization $P_x$ depend 
only on $V_0$, $k_Z$, and $\alpha$ for $\mathbf{B}||\mathbf{e}_x$ 
and not on $\alpha_y$ and $\alpha_z$ independently. 
The relevant energy scale of the polarization shown in  Fig.~\ref{step_pol}  
is given by $E_Z$, which defines the size of the ``gap'' [see Eq.~(\ref{energygap})]. 
Therefore, energies are given in units of $E_Z$ and wave vectors in units of $k_Z$. 
The same holds for the transmissions and polarizations shown further down (see 
Figs.~\ref{barrier_G} and \ref{barrier_P}). 
The parameters in Fig.~\ref{step_pol} are 
$V_0/E_Z=15$, and $\alpha/k_Z=2$, $2.5$, $3$, $5$. 
The energy offset is chosen such that $E_F/E_Z=0$ corresponds to the middle of the ``gap''. 
The parallel polarization $P_x$ gives the main contribution to the total polarization as 
the energy departs from the ``gap'', $P_x/P\rightarrow 1$. However, in
this region the total polarization is
negligible and within the ``gap'', the parallel component plays an
inferior role. 
The ratio of the two perpendicular polarizations
is given by $\left|P_z/P_y\right|=\alpha_y/\alpha_z$. Therefore, the
orthogonal polarization $\mathbf{P}_\perp=(0,P_y,P_z)$ can be rotated
within the $y$-$z$-plane by adjusting $\alpha_y$ and $\alpha_z$.

We next study the transmission current at a potential barrier of
height $V_0$ and width $2x_c$ [see Fig.~\ref{potential} b)]. This
situation might be more realistic than a simple potential step if one
thinks of further structuring by applying gates to the quantum wire.  
Fig.~\ref{barrier_G} shows the four components of the transmission as
a function of $E_F/E_Z$ for $\alpha/k_Z=2$, $2.5$, $3$, $V_0/E_Z=15$, 
and $k_Z x_c=1$. Again, the SOI affects the transmissions $t_{ss'}$
only via $\alpha$. Interestingly and in contrast to the potential
step, the $s$-flipping transmissions are degenerate,
$t_{+-}=t_{-+}$. This can be understood, if one considers the possible
$s$-flips at the two interfaces leading to an overall
$s$-flip.   
Labeling the left interface (1) and the right (2), one simply has 
to take the sum of the products of transmissions at each
interface and obtains
\begin{eqnarray}
t_{+-}&=&t_{++}(1)t_{+-}(2)+t_{+-}(1)t_{--}(2) \, , \nonumber \\
t_{-+}&=&t_{--}(1)t_{-+}(2)+t_{-+}(1)t_{++}(2) \, . \label{trans_flip}
\end{eqnarray}
An analysis of the potential step problem shows that the
$s$-conserving transmissions $t_{++}$ and $t_{--}$ are independent of
the sign of $V_0$ and the $s$-flipping transmissions just swap,
i.e. $t_{+-}(1)=t_{-+}(2)$ and $t_{-+}(1)=t_{+-}(2)$. This leads to
exactly the same values of $t_{+-}$ and $t_{-+}$ in
Eq.~(\ref{trans_flip}). The exponential suppression of $t_{++}(1)$ and
$t_{-+}(1)$ for energies within the ``gap'' does not affect this
behavior. The $s$-conserving transmissions $t_{++}$ and $t_{--}$ 
show an oscillatory behavior, which is well known from scattering
off a potential step at vanishing SOI. However, especially for
low energies, the amplitude strongly depends on $\alpha$. 
The $s$-flipping transmissions $t_{+-}$ and $t_{-+}$ oscillate as well.
The second peak of $t_{++}$, which lies in the ``energy gap'', is suppressed 
compared to $t_{--}$, since right-moving ($s=+$)-waves are exponentially 
damped in the barrier region and therefore, as shown in 
Ref.~\onlinecite{streda}, $t_{--}$ is the dominant component at each 
interface in this energy range. 

Fig.~\ref{barrier_P} shows $P$ and $P_x/P$ for
the same parameters as in Fig.~\ref{barrier_G}. 
and $\alpha/k_Z=2$, $2.5$, $3$. 
Similarly to the potential step
case, $P=|\mathbf{P}|$ and $P_x$ only depend on $\alpha$ and not on
$\alpha_y$ and $\alpha_z$ independently. 
Surprisingly, the polarization now has a sizable value in an
energy interval much bigger than the ``gap'', which just goes from
$-E_Z$ to $E_Z$ (see the arrows in Fig.~\ref{barrier_P}). 
This behavior must be contrasted to the polarization in the case of a potential 
step as shown in Fig.~\ref{step_pol} and first introduced in Ref.~\onlinecite{streda}.
It can be traced back to the energy dependence of $t_{+-}$
and $t_{-+}$ shown in Fig.~\ref{barrier_G}. Both have finite weight 
well beyond the ``energy gap''. This might be due to interference effects 
of transmitted and reflected waves in the barrier region.

\section{Lattice model}

In a next step, we are aiming at constructing a tight-binding 
lattice model which in appropriate parameter regimes shows similar
physics as our continuum model. This will put us in a position to 
study the effect of electron-electron interaction neglected so far 
using the functional renormalization group method.\cite{funRG} 
In 1D wires the two-particle interaction is known to strongly alter the
low-energy physics of many-body systems leading to so called
Luttinger liquid behavior.\cite{KS} It can be expected that the
interplay of the SOI effects discussed above and correlation 
effects leads to interesting physics.
\begin{figure*}[t]
\begin{center}
\includegraphics[width=0.3\textwidth,clip]{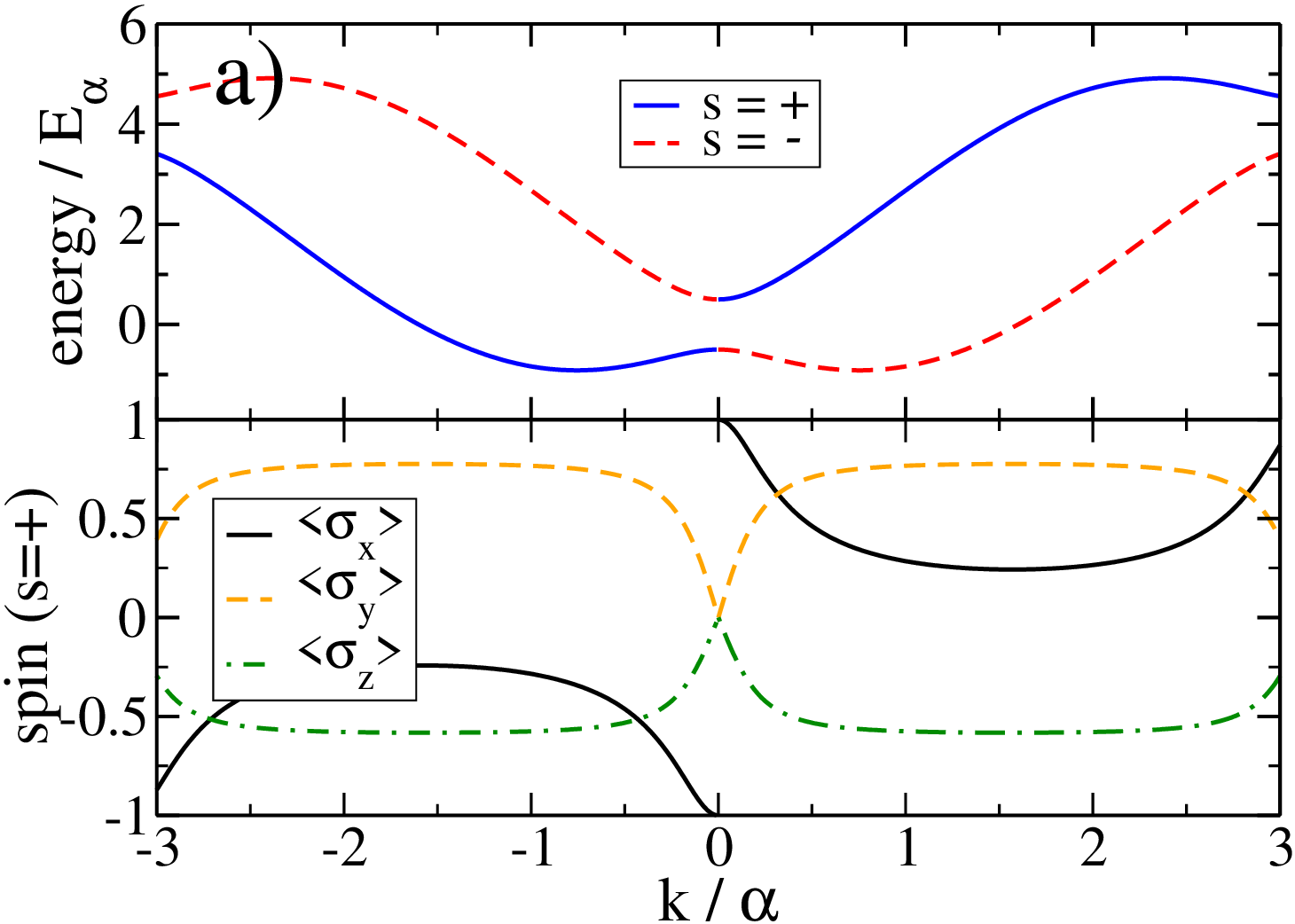}\hspace{0.5cm}
\includegraphics[width=0.3\textwidth,clip]{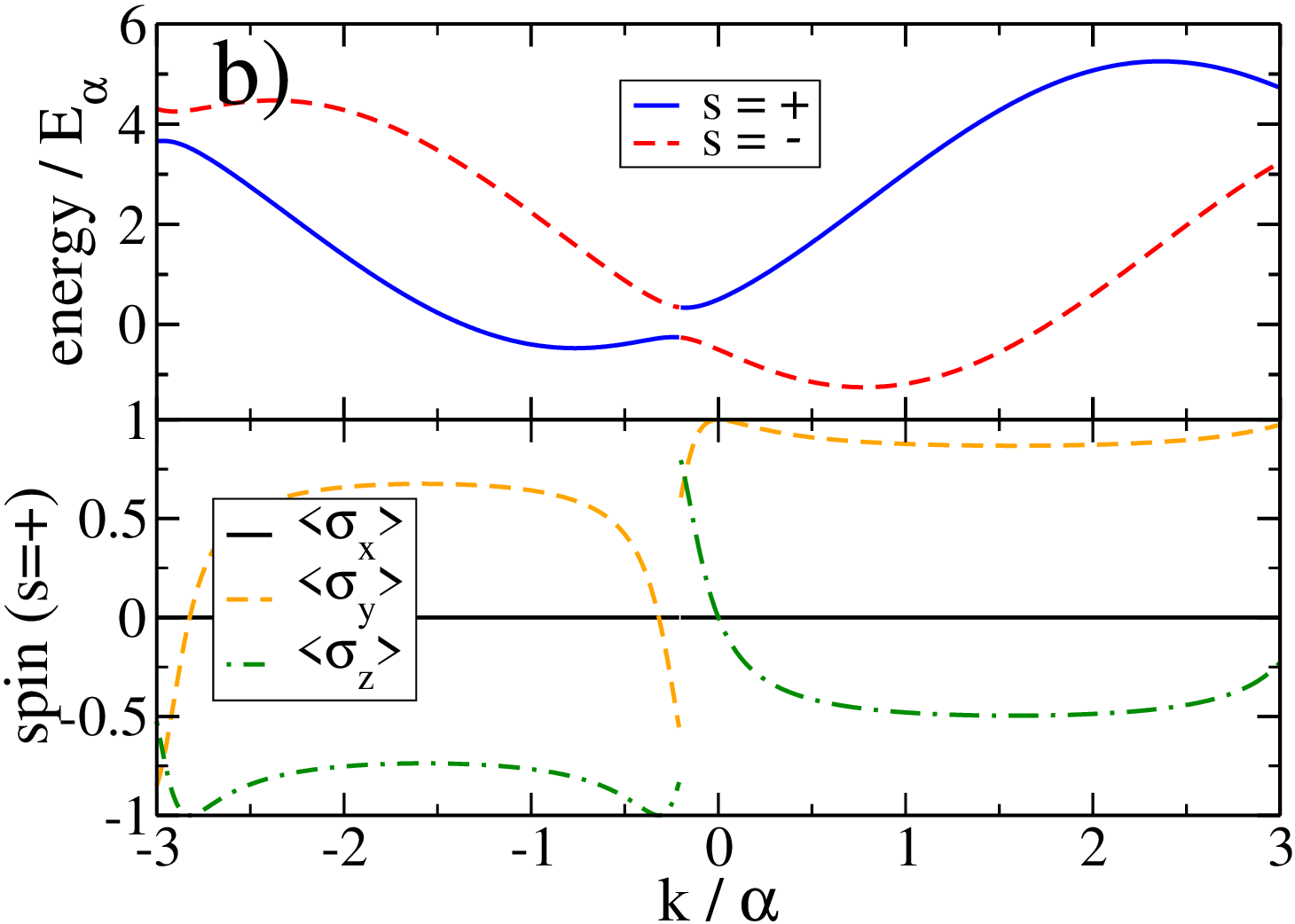}\hspace{0.5cm}
\includegraphics[width=0.3\textwidth,clip]{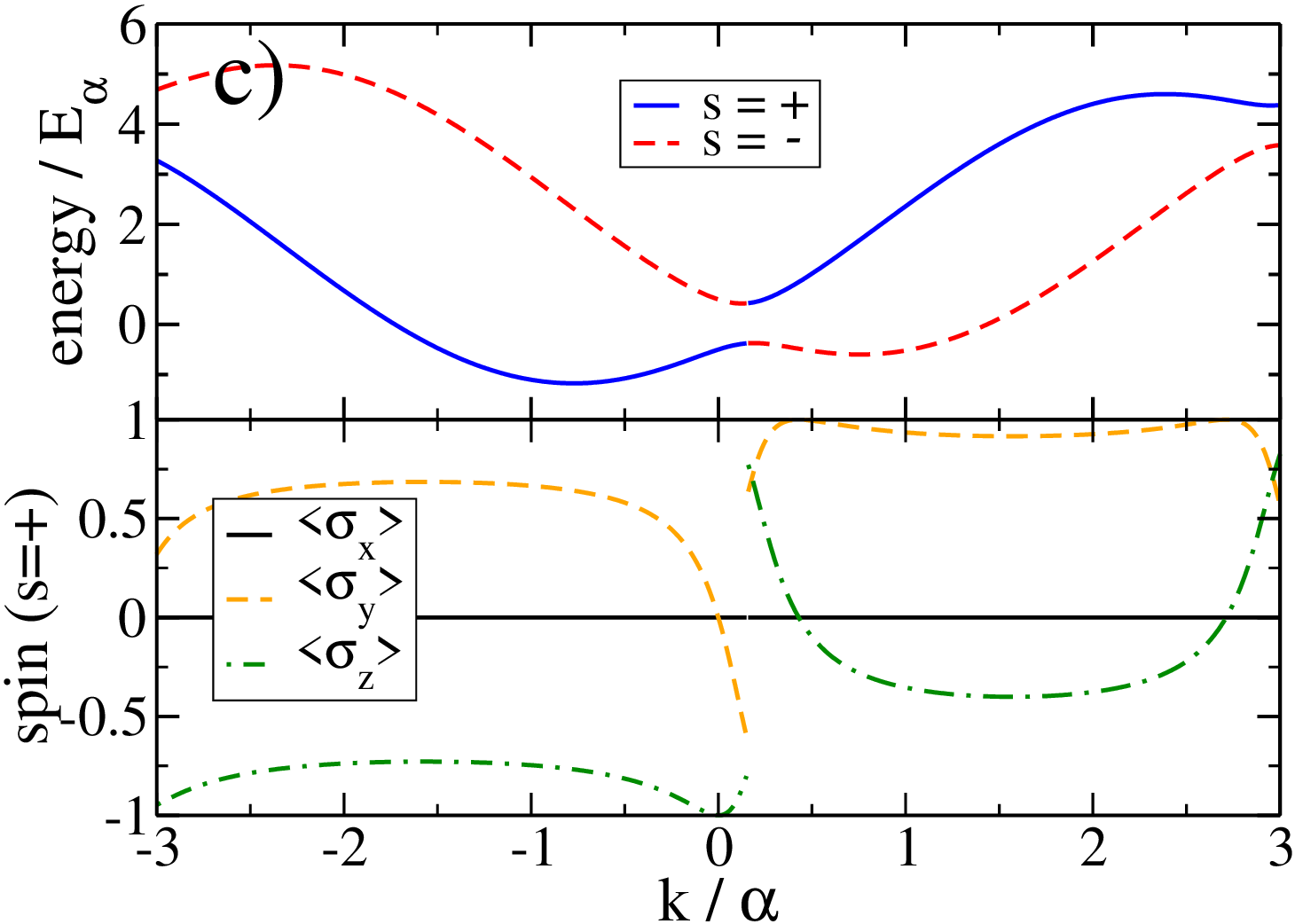}
\end{center}
\vspace{-0.6cm}
\caption[]{(Color online) Lattice dispersion and spin expectation
  values on the ($s=+$)-branch  for a magnetic field in a) $x$- , b) $y$- and c)
  $z$-direction for $t/\alpha=1$, $\alpha_y/\alpha=-0.6$,
  $\alpha_z/\alpha=-0.8$, $k_Z/\alpha=0.5$. 
  The spin on the ($s=-$)-branch points
  in the opposite direction, i.e.~$\langle \sigma_i\rangle_{k,s}=
  -\langle \sigma_i\rangle_{k,-s}$.
For
  $\left|k-k_0\right|<\pi/2$ one obtains exactly the same behavior as
  in the continuum case.} 
\label{lattice_spindisp}
\end{figure*}
The SOI can be modeled by spin-flip hopping terms with 
amplitude $\alpha_y$ and $\alpha_z$ in a usual tight-binding 
model.\cite{mireles} 

We start with a representation of the
Hamiltonian in terms of Wannier states $|j,\sigma\rangle$ with
$j\in\mathbb{Z}$ labeling the lattice site and
$\sigma=\uparrow,\downarrow$ labeling the spin. 
The spin quantization is chosen along the $z$-direction. 
With $c_{j,\sigma}^\dagger$ being the creation operator of an electron
at site $j$ with spin $\sigma$, the lattice model Hamiltonian for an
arbitrary magnetic field
$\mathbf{B}=B\left(\sin\theta\cos\varphi,\sin\theta\sin\varphi,\cos\theta\right)$
can be written as
\begin{equation}
H=H_0+H_\mathrm{pot}+H_R+H_Z \, ,
\end{equation}
with the free part
\begin{equation}
H_0=\epsilon\sum_{j,\sigma} c_{j,\sigma}^\dagger
  c_{j,\sigma}-t\sum_{j,\sigma} 
\left(c_{j+1,\sigma}^\dagger c_{j,\sigma} + c_{j,\sigma}^\dagger  
c_{j+1,\sigma}\right) \, ,
\end{equation}
containing the on-site energy and the conventional (spin-conserving)
hopping, external potential (due to e.g.~nano-device structuring)
\begin{equation}
H_\mathrm{pot}=\sum_{j,\sigma} V_{j,\sigma} 
c_{j,\sigma}^\dagger c_{j,\sigma} \, ,
\end{equation}
the spin-flip (Rashba) hopping terms
\begin{eqnarray}
H_R&=&-\alpha_z\sum_{j,\sigma,\sigma'}\left( c_{j+1,\sigma}^\dagger
\left(i\sigma_y\right)_{\sigma,\sigma'} c_{j,\sigma'} + \mbox{H.c.}
\right)
\\ \nonumber
&&+\alpha_y\sum_{j,\sigma,\sigma'}\left( c_{j+1,\sigma}^\dagger
\left(i\sigma_z\right)_{\sigma,\sigma'} c_{j,\sigma'} + \mbox{H.c.} \right) \, ,
\end{eqnarray}
and the Zeeman term 
\begin{eqnarray}
H_Z&=& 2k_Z^2 \sum_{j,\sigma,\sigma'}c_{j,\sigma}^\dagger\left[
  \left(\sigma_x\right) 
_{\sigma,\sigma'}\sin\theta\cos\varphi \right. \\ \nonumber
&& + \left. \left(\sigma_y\right) 
_{\sigma,\sigma'}\sin\theta\sin\varphi+\left(\sigma_z\right) 
_{\sigma,\sigma'}\cos\theta \right]  c_{j,\sigma'} \, .
\end{eqnarray}
We show the analogy to the continuum case suppressing 
$H_\mathrm{pot}$ and take as an ansatz for the corresponding eigenstates
\begin{equation}
|k,s\rangle =\sum\limits_{j,\sigma}a_{\sigma}^s(k)
e^{ikj}|j,\sigma\rangle \,  .\label{ansatz2} 
\end{equation} 
This leads to the eigenenergies
\begin{equation}
E^{(s)}(k)=\epsilon-2t\cos k + 2s\sgn(k-k_0)\sqrt{D(k)} \, ,\label{eigenvalue2}
\end{equation}
with 
\begin{equation}
k_0=\arcsin\left[-k_Z^2
  (\alpha_y\cos\theta-\alpha_z\sin\theta\sin\varphi)
/(\alpha_y^2+\alpha_z^2)\right] 
\end{equation}
and
\begin{eqnarray}
D(k)&=&(\alpha_y^2+\alpha_z^2)\sin^2 k +k_Z^4\\ 
\nonumber
&& + 2k_Z^2\sin k\left(\alpha_y \cos\theta -\alpha_z\sin\theta\sin\varphi
\right)\, .
\end{eqnarray}
Eq.~(\ref{eigenvalue2}) has almost the same form as the continuum 
version Eq.~(\ref{eigenvalue}). In fact, choosing the on-site energy
$\epsilon=2t$, which corresponds just to an overall energy shift, and 
substituting $\cos k$ by $1-k^2/2$ and $\sin k$ by $k$, which is valid
for sufficiently small $|k|$, we get exactly the same form. Note however
that, in contrast to the continuum case, $\alpha_y,\alpha_z$ and
$k_Z^2$ now have the unit of energy, but since $c_k^{(s)}$, defined 
in Eq.~(\ref{cks}) is dimensionless, all formulas remain valid.
We choose for the eigenstates Eq.~(\ref{ansatz2}) 
$a_\downarrow^s(k)=1$ and obtain $a_\uparrow^s(k)=c_k^{(s)}$ 
\begin{widetext}
\begin{equation}
\mbox{with}\quad c_k^{(s)}= \frac{-i\alpha_z\sin k-k_Z^2
  e^{-i\varphi}\sin\theta}{\alpha_y\sin k + k_Z^2 \cos\theta - s\sgn
  (k-k_0)\sqrt{(\alpha_y^2+\alpha_z^2)\sin^2 k+2k_Z^2\sin
    k(\alpha_y\cos\theta - \alpha_z\sin\theta\sin\varphi)+k_Z^4}},\label{cks}
\end{equation}
\end{widetext}
and the spin expectation values have exactly the continuum form
\begin{equation}
\langle \sigma_x + i\sigma_y\rangle_{k,s} =
2\frac{\left(c_k^{(s)}\right)^*}{1+\left|c_k^{(s)}\right|^2}, \quad
\langle \sigma_z \rangle_{k,s} =
\frac{-1+\left|c_k^{(s)}\right|^2}{1+\left|c_k^{(s)}\right|^2}
\, . \label{spin2}
\end{equation}

The energy dispersions and the spin expectation values for magnetic
fields in $x$-, $y$-, and $z$-direction are shown in
Fig.~\ref{lattice_spindisp}. 
Besides the cosine-like structure, which becomes especially relevant
near the band edges, the dispersion and spin
expectation values have the same shape as in the continuum model. 
A direct comparison of Fig.~\ref{lattice_spindisp} and
Fig.~\ref{spindisp} 
shows that our lattice model reproduces the low energy physics, i.e.~for
$\left|k-k_0\right|<\pi/2$, observed in the continuum. 
As above we only show the spin expectation values on the ($s=+$)-branch. 
The spin on the ($s=-$)-branch points
  in the opposite direction, i.e.~$\langle \sigma_i\rangle_{k,s}=
  -\langle \sigma_i\rangle_{k,-s}$. The direct relation 
between the dispersion and the spin expectation values for energies of
the order of the ``gap'' is the essential feature leading to the
remarkable scattering properties of the continuum model (and
eventually a spin polarized
conductance) at steps and barriers. One can thus expect similar transport
characteristics to be realized in the lattice model. A detailed
discussion of this and in particular
the effect of the electron-electron interaction
on transport will be the topic of an upcoming
publication.\cite{interactionpaper}

\section{Conclusions} 

We have investigated the dispersion and spin expectation
values of a 1D electron system with SOI as well as arbitrary 
magnetic field, and have shown that an additional SOI term 
resulting from the lateral confinement of a 2DEG to a 1D wire 
geometry leads to a rotation of the spin out of the 2D plane.
For the case of a magnetic field parallel to the 
quantum wire, the transmission and polarization of a linear response 
current at a potential step as well as at a potential barrier were
studied. For the latter, we observed an extended energy range, where 
significant spin polarization can be achieved. We showed that this
spin polarization can be rotated out of the plane of the 2DEG
arbitrarily by adjusting the SOI constants $\alpha_y$
and $\alpha_z$. The potential barrier describes a setup which can
experimentally be achieved by adding further gates to the wire geometry.
We then constructed a lattice model which shows the same low energy
physics as the continuum model. This lattice model now enables us to 
investigate the interplay of SOI and Coulomb 
interaction in quantum wires with potential steps and barriers using
the functional renormalization group method.\cite{interactionpaper,funRG}

\section*{Acknowledgments}
We thank J.~Sinova and U.~Z\"ulicke for useful discussions. 
This work was supported by the the Deutsche 
Forschungsgemeinschaft via SFB 602.


\begin{thebibliography}{*}
\bibitem {review} I.~\v{Z}uti\'{c}, J.~Fabian, and S.~Das Sarma,
  Rev.~Mod.~Phys.~{\bf 76}, 323 (2004).
\bibitem{winkler} R.~Winkler, \textit{Spin-Orbit Coupling Effects in Two-Dimensional Electron and Hole Systems}, Springer-Verlag, Berlin (2003). 
\bibitem{rashba1} E.I.~Rashba, Physica E {\bf 34}, 31 (2006).
\bibitem{footnote1} The SOI coupling coefficients $\alpha_i$ are assumed to 
be independent from the coordinates $j\neq i$,
i.e. $\partial_i\partial_j V=0$, 
with $i,j=x,y,z$ and $i\neq j$.
\bibitem{datta} S.~Datta and B.~Das, Appl.~Phys.~Lett.~{\bf 56}, 7 (1990).
\bibitem{moroz} A.V.~Moroz and C.H.W.~Barnes, Phys.~Rev.~B {\bf 60},
  14272 (1999).
\bibitem{mireles} F.~Mireles and G.~Kirczenow, Phys.~Rev.~B {\bf 64}, 24426 (2001).
\bibitem{uli} M.~Governale and U.~Z\"ulicke, Phys.~Rev.~B {\bf 66},
  07331 (2002); Solid State Com.~{\bf
    131}, 581 (2004).
\bibitem{streda} P.~Str\v{e}da and P.~ S\v{e}ba, Phys.\ Rev.\ Lett.\ {\bf
    90}, 256601 (2003).
\bibitem{pereira} R.G.~Pereira and E.~Miranda,  Phys.~Rev.~B {\bf 71},
  085318 (2005). 
\bibitem{serra} L.~Serra, D.~S\'{a}nchez, and R.~L\'{o}pez,
  Phys.~Rev.~B {\bf 72}, 235309 (2005).
\bibitem{zhang} S.~Zhang, R.~Liang, E.~Zhang, L.~Zhang, and Y.~Liu,
  Phys.~Rev.~B {\bf 73}, 155316 (2006). 
\bibitem{interactionpaper} J.E.~Birkholz and V.~Meden, in preparation.  
\bibitem{funRG} S.~Andergassen, T.~Enss, V.~Meden, 
W.~Metzner, U.~Schollw\"ock, and K.~Sch\"onhammer, 
Phys.\ Rev.\ B {\bf 73}, 045125 (2006).
\bibitem{molenkamp} L.W.~Molenkamp, G.~Schmidt, and G.E.W.~Bauer, Phys.~Rev.~B {\bf 64}, 121202 (2001).
\bibitem{KS}
For a recent review see
K.~Sch\"onhammer in {\it Interacting Electrons in Low 
 Dimensions,} Ed.: D.~Baeriswyl, Kluwer Academic Publishers 
 (2005).
\end{thebibliography}
\end{document}